\documentclass[twoside,twocolumn,9pt]{article}
\usepackage{extsizes}
\usepackage[super,sort&compress,comma]{natbib} 
\usepackage[version=3]{mhchem}
\usepackage[left=1.5cm, right=1.5cm, top=1.785cm, bottom=2.0cm]{geometry}
\usepackage{balance}
\usepackage{mathptmx}
\usepackage{sectsty}
\usepackage{graphicx} 
\usepackage{lastpage}
\usepackage{gensymb}
\usepackage[format=plain,justification=justified,singlelinecheck=false,font={stretch=1.125,small,sf},labelfont=bf,labelsep=space]{caption}
\usepackage{float}
\usepackage{fancyhdr}
\usepackage{fnpos}
\usepackage[english]{babel}
\addto{\captionsenglish}{%
  \renewcommand{\refname}{Notes and references}
}
\usepackage{array}
\usepackage{droidsans}
\usepackage{charter}
\usepackage[T1]{fontenc}
\usepackage[usenames,dvipsnames]{xcolor}
\usepackage{setspace}
\usepackage[compact]{titlesec}
\usepackage{hyperref}

\usepackage{epstopdf}

\definecolor{cream}{RGB}{222,217,201}

\begin{document}

\pagestyle{fancy}
\thispagestyle{plain}
\fancypagestyle{plain}{
\renewcommand{\headrulewidth}{0pt}
}

\makeFNbottom
\makeatletter
\renewcommand\LARGE{\@setfontsize\LARGE{15pt}{17}}
\renewcommand\Large{\@setfontsize\Large{12pt}{14}}
\renewcommand\large{\@setfontsize\large{10pt}{12}}
\renewcommand\footnotesize{\@setfontsize\footnotesize{7pt}{10}}
\makeatother

\renewcommand{\thefootnote}{\fnsymbol{footnote}}
\renewcommand\footnoterule{\vspace*{1pt}%
\color{cream}\hrule width 3.5in height 0.4pt \color{black}\vspace*{5pt}} 
\setcounter{secnumdepth}{5}

\makeatletter 
\renewcommand\@biblabel[1]{#1}            
\renewcommand\@makefntext[1]%
{\noindent\makebox[0pt][r]{\@thefnmark\,}#1}
\makeatother 
\renewcommand{\figurename}{\small{Fig.}~}
\sectionfont{\sffamily\Large}
\subsectionfont{\normalsize}
\subsubsectionfont{\bf}
\setstretch{1.125} 
\setlength{\skip\footins}{0.8cm}
\setlength{\footnotesep}{0.25cm}
\setlength{\jot}{10pt}
\titlespacing*{\section}{0pt}{4pt}{4pt}
\titlespacing*{\subsection}{0pt}{15pt}{1pt}

\fancyfoot{}
\fancyfoot[LO,RE]{\vspace{-7.1pt}\includegraphics[height=9pt]{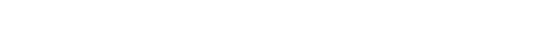}}
\fancyfoot[CO]{\vspace{-7.1pt}\hspace{13.2cm}\includegraphics{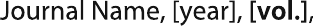}}
\fancyfoot[CE]{\vspace{-7.2pt}\hspace{-14.2cm}\includegraphics{head_foot/RF}}
\fancyfoot[RO]{\footnotesize{\sffamily{1--\pageref{LastPage} ~\textbar  \hspace{2pt}\thepage}}}
\fancyfoot[LE]{\footnotesize{\sffamily{\thepage~\textbar\hspace{3.45cm} 1--\pageref{LastPage}}}}
\fancyhead{}
\renewcommand{\headrulewidth}{0pt} 
\renewcommand{\footrulewidth}{0pt}
\setlength{\arrayrulewidth}{1pt}
\setlength{\columnsep}{6.5mm}
\setlength\bibsep{1pt}

\makeatletter 
\newlength{\figrulesep} 
\setlength{\figrulesep}{0.5\textfloatsep} 

\newcommand{\topfigrule}{\vspace*{-1pt}%
\noindent{\color{cream}\rule[-\figrulesep]{\columnwidth}{1.5pt}} }

\newcommand{\botfigrule}{\vspace*{-2pt}%
\noindent{\color{cream}\rule[\figrulesep]{\columnwidth}{1.5pt}} }

\newcommand{\dblfigrule}{\vspace*{-1pt}%
\noindent{\color{cream}\rule[-\figrulesep]{\textwidth}{1.5pt}} }

\makeatother

\twocolumn[
  \begin{@twocolumnfalse}
{\includegraphics[height=30pt]{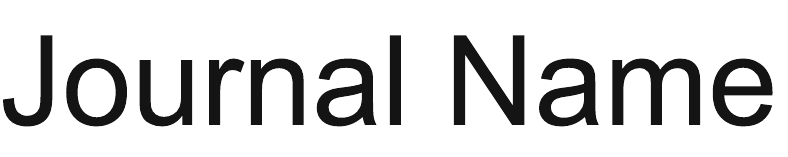}\hfill\raisebox{0pt}[0pt][0pt]{\includegraphics[height=55pt]{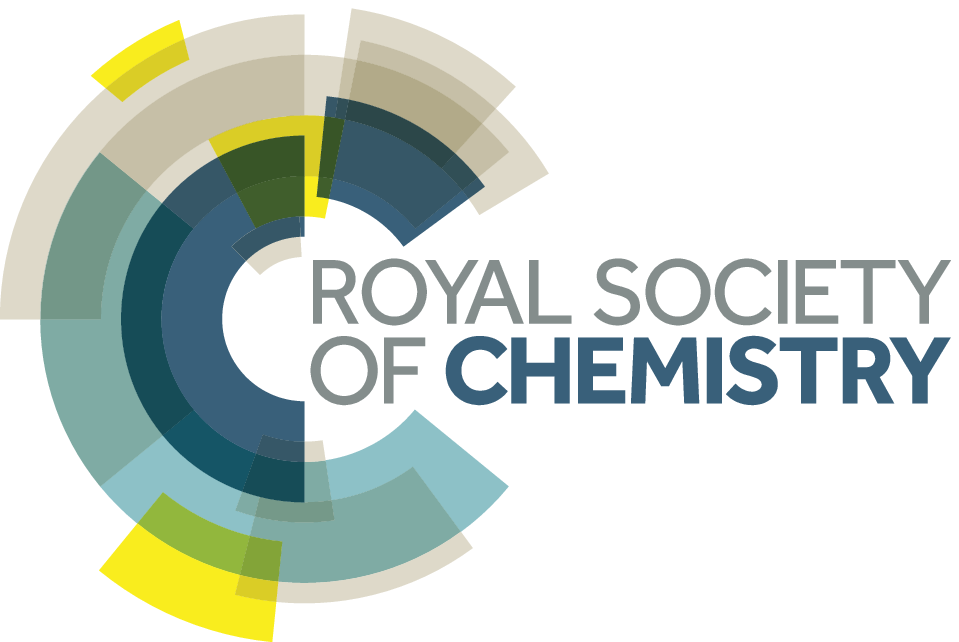}}\\[1ex]
\includegraphics[width=18.5cm]{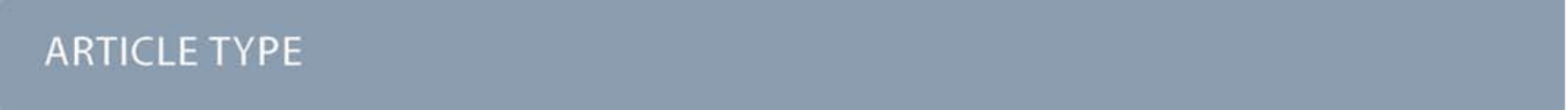}}\par
\vspace{1em}
\sffamily
\begin{tabular}{m{4.5cm} p{13.5cm} }

\includegraphics{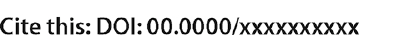} & \noindent\LARGE{\textbf{High-throughput mechanophenotyping of multicellular spheroids using a microfluidic micropipette aspiration chip$^\dag$}} \\
\vspace{0.3cm} & \vspace{0.3cm} \\

 & \noindent\large{Ruben C. Boot,\textit{$^{a}$} Alessio Roscani,\textit{$^{a}$} Lennard van Buren,\textit{$^{b}$} Samadarshi Maity,\textit{$^{a}$} Gijsje H. Koenderink\textit{$^{b}$} and Pouyan E. Boukany\textit{$^{a}$}} \\

\includegraphics{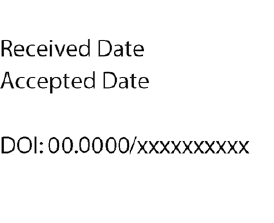} & \noindent\normalsize{Cell spheroids are \textit{in vitro} multicellular model systems that mimic the crowded micro-environment of biological tissues. Their mechanical characterization can provide valuable insights in how single-cell mechanics and cell-cell interactions control tissue mechanics and self-organization. However, most measurement techniques are limited to probing one spheroid at a time, require specialized equipment and are difficult to handle. Here, we developed a microfluidic chip that follows the concept of glass capillary micropipette aspiration in order to quantify the viscoelastic behavior of spheroids in an easy-to-handle, high-throughput manner. Spheroids are loaded in parallel pockets via a gentle flow, after which spheroid tongues are aspirated into adjacent aspiration channels using hydrostatic pressure. After each experiment, the spheroids are easily removed from the chip by reversing the pressure and new spheroids can be injected. The presence of multiple pockets with a uniform aspiration pressure, combined with the ease to conduct successive experiments, allows for a high throughput of tens of spheroids per day. We demonstrate that the chip provides accurate deformation data when working at different aspiration pressures. Lastly, we measure the viscoelastic properties of spheroids made of different cell lines and show how these are consistent with previous studies using established experimental techniques. In summary, our chip provides a high-throughput way to measure the viscoelastic deformation behavior of cell spheroids, in order to mechanophenotype different tissue types and examine the link between cell-intrinsic properties and overall tissue behavior.} \\

\end{tabular}

 \end{@twocolumnfalse} \vspace{0.6cm}

  ]

\renewcommand*\rmdefault{bch}\normalfont\upshape
\rmfamily
\section*{}
\vspace{-1cm}


\footnotetext{\textit{$^{a}$~Department of Chemical Engineering, Delft University of Technology, Delft, The Netherlands. E-mail: p.e.boukany@tudelft.nl}}
\footnotetext{\textit{$^{b}$~Department of Bionanoscience, Kavli Institute of Nanoscience Delft, Delft University of Technology, Delft, The Netherlands }}

\footnotetext{\dag~Electronic Supplementary Information (ESI) available: [details of any supplementary information available should be included here]. See DOI: 00.0000/00000000.}


\section*{Introduction}
Physical properties like cellular mechanics are of undeniable importance in physiological processes such as morphogenesis\cite{Hahn2009}, tissue remodeling\cite{Mammoto2010}, wound-healing\cite{Brugues2014} and cancer growth\cite{Nia2016,Nia2020}. During these events, cells are collectively confined, squeezed, pushed or pulled upon, affecting their self-organization in time and space. The overall mechanical response to these forces, termed tissue mechanics, will shape the resulting tissue morphology. This response depends on the properties of the single cells alongside the interplay between cells across multiple length scales. \cite{Jakab2008, Foty1996} While the mechanical deformation of single cells mostly depends on their cytoskeleton, plasma membrane and nuclear stiffness, tissue mechanics are defined through the linkage between cell adhesion molecules, the cytoskeleton and the extracellular environment. \cite{Marturano2013,Schiele2015}

Cell spheroids have become a popular \textit{in vitro} model to study tissue mechanics, as they replicate both the multicellular nature and three-dimensional (3D) micro-environment of \textit{in vivo} tissues. \cite{Gonzalez-Rodriguez2012} These spherical aggregates are made from immortalized cell lines or primary cells that adhere to each other and collectively round up. The resulting spheroid morphology and internal cell arrangement is defined by the interplay between cell-cell adhesion and cortical tension. \cite{Manning2010,Foty1996,Brodland2002} Probing spheroids with relevant physical forces therefore increases insight in how tissue composition and resulting mechanics relate to tissue sorting, cellular mechanosensing and cell invasion. \cite{Jakab2008,Dufour2010,Han2020} 

Spheroid mechanics have been quantified using various techniques, probing either from within or without. \cite{Boot2021} For example, hydrogel mechanosensors give information on the spatial distribution of mechanical stress within spheroids. \cite{Mok2020,Dolega2017} Cavitation rheology probes the internal elasticity and tissue interfacial tension by inducing a spherical cavity in the spheroid with a needle. \cite{Blumlein2017} From without, the elastic modulus has been quantified by squeezing the spheroid between two "chopsticks" termed microtweezers. \cite{Jaiswal2017} Atomic force microscopy (AFM) determines the viscoelastic response of a spheroid by indenting the surface with a nano-probe \cite{Dolega2021,Vyas2019}, while tissue surface tensiometry (TST) squeezes the spheroid between two plates \cite{Manning2010,Foty1996,Schotz2008,Norotte2008,Mgharbel2009} and micropipette aspiration (MPA) aspirates a spheroid tongue in a glass capillary to look at the viscoelastic creep response. \cite{Dufour2010,Guevorkian2011,Yousafzai2022,Yadav2022} Additionally, TST and MPA quantify a tissue surface tension, for which the analogy is made between round spheroids and liquid droplets. \cite{Manning2010,Foty1994,Dufour2010} This surface tension is directly related to tissue sorting, tissue spreading and energetic constraints on the size of spheroids. \cite{Foty1996,Ryan2001,Yousafzai2022}

However, available techniques to quantify spheroid mechanical parameters such as the elastic modulus $E$, viscosity $\eta$ and surface tension $\gamma$ have a limited throughput. First, techniques such as AFM, microtweezers, TST or MPA only probe one spheroid at a time. Second, the handling of nano-cantilevers, small tweezers or glass microcapillaries is a delicate, difficult and time-consuming task. The resulting low throughput is found in the previously mentioned studies probing spheroid mechanics, as these usually report a data set that ranges between a total number of $\sim$5 to 30 probed spheroids. As the technical challenges form a bottleneck on the size of data sets, it is difficult to quantify differences between various spheroid models using present techniques. Given the fact that biological variability tends to be rather large, mechanical phenotyping, for instance to compare different cancer types, or in-depth studies of the role of cytoskeletal components or specific (cancer) biomarkers in overall tissue behavior require an assay with higher throughput than what is currently available.

Microfluidic devices are widely used to measure the mechanical properties of single cells at high throughput. \cite{Truongvo2017,Mak2013, Mak2013a,Hou2009,Davidson2019,Lee2015,Au2016,Cognart2020,Raj2017,Adamo2012,Chang2019} Here, cell deformability is examined by letting large numbers of cells flow or migrate through narrow channels or micro-pillars. The chip's defined geometries are easily replicated into new chips, making this a highly reproducible set-up. Besides overall cell deformability, these devices are able to quantify more specific mechanical parameters such as both the cell's and nuclear elastic modulus and viscosity. For example, the design of a microfluidic array where single cells land in individual pockets and are aspirated via a pressure gradient enables high-throughput micropipette aspiration. \cite{Davidson2019,Lee2015} However, applying the same principles to study mechanics of spheroids or tissues, which requires microfluidic chips with larger channel dimensions, has remained unaddressed. While microfluidic devices exist to examine spheroid growth and for functional assessment of drugs\cite{Sabhachandani2016,Ruppen2014}, a high-throughput microfluidic chip to study spheroid mechanics in parallel and with high reproducibility does not exist to date.

Inspired by the microfluidic micropipette array for single cells \cite{Davidson2019,Lee2015}, we have designed a microfluidic device to perform MPA on multiple spheroids in parallel, thereby drastically increasing the throughput. After each measurement, spheroids can be easily removed from the device by reversing the flow and aspirating them at the inlet, allowing for multiple experiments per chip. As the device is made from a mold, each chip has the exact same dimensions for the micropipette channels, which is much harder to obtain when pulling glass micropipettes for traditional MPA. The aspiration pressure is precise and easily controlled for each measurement by using hydrostatic pressure. With a custom-made Python script for automated image analysis, the creep length of aspirated spheroid tongues can be analyzed to derive the viscoelastic response of the spheroids. Our device can aspirate 8 spheroids in parallel per measurement for multiple runs per day, allowing for much larger data sets while providing the same information as traditional glass micropipette aspiration. Additionally, we show that our microfluidic device is sensitive enough to pick up mechanical differences between different spheroid models, making it a suitable device to mechanically phenotype different cellular systems. 

\section*{Materials and methods}
\subsection*{Cell culture}
Human embryonic kidney 293 (HEK293T) cells were generously provided by the group of Dimphna Meijer (Department of Bionanoscience, Delft University of Technology). They were kept in Dulbecco's Modified Eagle Medium High Glucose (DMEM, Sigma) containing 4.5 g/L glucose, L-glutamine but without sodium pyruvate, and supplemented with 10\% Fetal Bovine Serum (FBS, Sigma) and 1\% Antibiotic-Antimycotic solution (Gibco). 

NIH3T3 embryonic mouse fibroblasts (ATCC CRL-1658) were kept in Dulbecco's Modified Eagle Medium High Glucose (DMEM, Sigma) containing 4.5 g/L glucose, L-glutamine but without sodium pyruvate, and supplemented with 10\% Newborn Calf Serum (NCS, Sigma) and 1\% Antibiotic-Antimycotic solution (Gibco).

Human mammary MCF10A cells (ATCC CRL-10317) were cultured in DMEM/F12 1:1 medium (Gibco) supplemented with 5\% horse serum (Gibco), 0.5 \textmu g/mL hydrocortisone (Sigma), 20 ng/mL human epidermal growth factor (hEGF) (Peprotech), 100 ng/mL cholera toxin (Sigma), 10 \textmu g/mL insulin (Human Recombinant Zinc, Gibco) and 1\% Penicillin-Streptomycin 100x solution (VWR Life Science).

All cells were incubated at 37 \degree C with 5\% CO${_2}$ and subcultured at least twice a week. 

\subsection*{Spheroid fabrication}
Spheroids were generated using a custom-designed microfabricated microwell array platform (which is available at: https://github.com/RubenBoot/HighThroughput\_Spheroid\_MPA
/blob/main/SpheroidMicrowellArray.dwg), inspired by work from Minglin Ma's lab.\cite{Song2016} Following their protocol, two microwell array platforms were created using standard soft lithography at the Kavli Nanolab Delft to allow for the creation of spheroids with different diameters. Using SU-8 2150 photoresist (Kayaku Advanced Materials) and a \textmu MLA laserwriter (Heidelberg Instruments), the master wafers were designed to have an array of circular microposts. The first wafer had posts with a diameter of 200 \textmu m and height of 220\textpm 20 \textmu m, while the second wafer had posts with a diameter of 280 \textmu m and a height of 300\textpm 30 \textmu m. The master wafers were coated with trichloro(1H,1H,2H,2H-perfluorooctyl)silane (Sigma-Aldrich) to allow for easy demolding. Microwell arrays were molded from the wafers using polydimethylsiloxane (PDMS) (Sylgard 184, Dow Corning) and curing agent at a mixing ratio of 10:1 (w/w). The arrays were placed in a 12-well cell culture plate (Thermo Fisher Scientific) using rubber glue (Reprorubber), and then sterilized by thoroughly washing with ethanol and leaving under UV light overnight.

Before seeding cells, the arrays were coated with 1\% (w/v) Pluronic\textsuperscript{\textregistered} F127 (Sigma-Aldrich) solution to prevent cell adhesion to the PDMS. The pluronic solution was removed from the well after 45 minutes of incubating. A cell suspension with a concentration of \textpm 1 x 10\textsuperscript{6} cells in matching cell media, obtained through trypsinization, was deposited in the well with the coated array in order to form spheroids. It is important to note that the resulting spheroid dimensions not only depend on the chosen cell concentration but also on the duration of culture, cell adhesion and the cell type-specific proliferation rate. After deposition, cells divide over the microwells and settle at the bottom due to gravity, where they aggregate into spheroids overnight (Fig. S1\dag). The spheroids were cultured in the wells for either 2 or 3 days before aspiration experiments, changing the media every day. On the day of the experiment, spheroids were gently washed out of the microwells using the same media and brought into suspension. 

\subsection*{Design and fabrication of the microfluidic chip}
\begin{figure}[h]
\centering
\includegraphics[height=12.2cm]{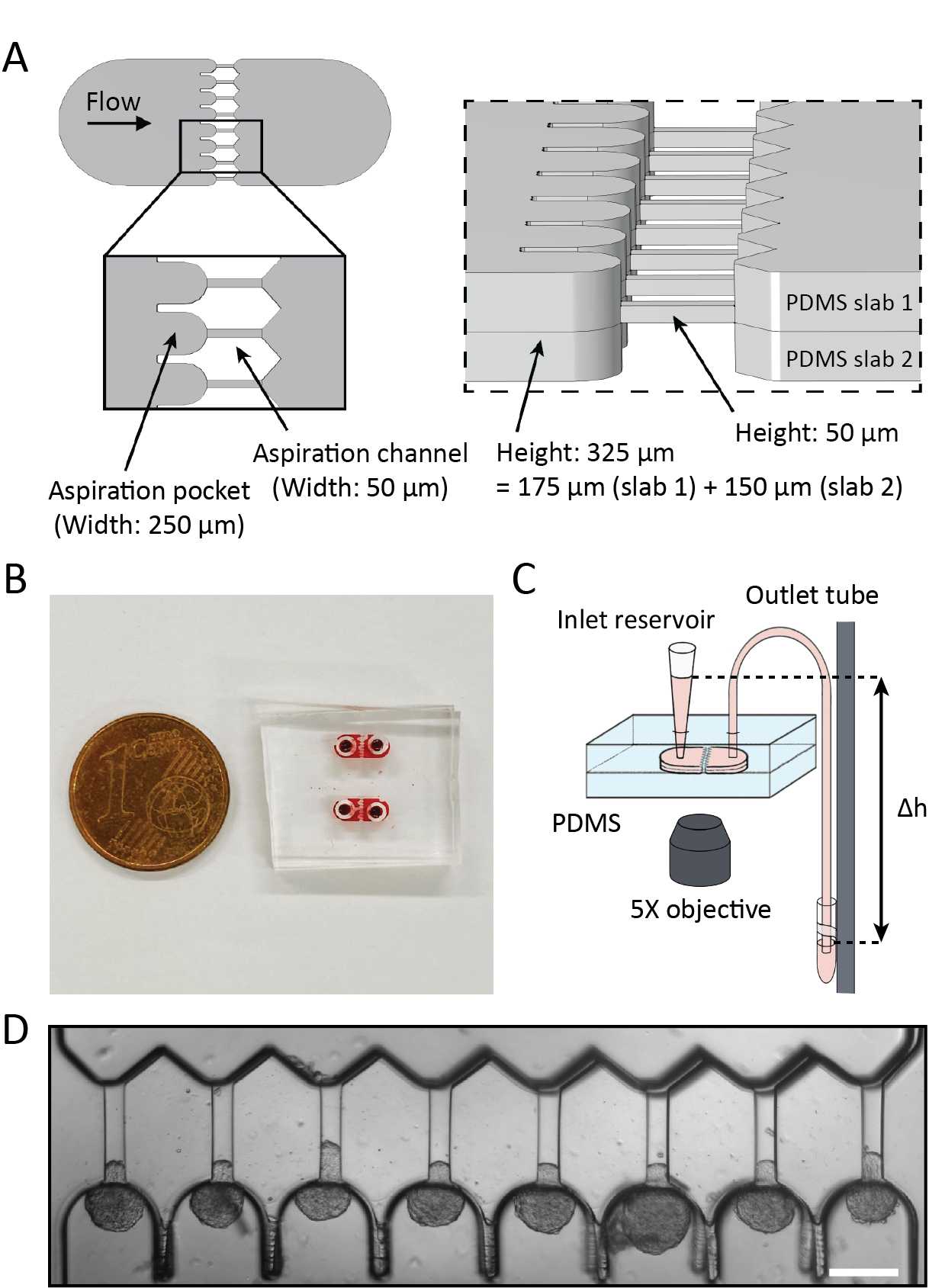}
\caption{\label{Fig1} Overview of the microfluidic chip. (A) 3D schematic showing (left) the top view of the design and a close-up on the pockets and aspiration channels, and (right) a tilted side view showing the heights of the two separate PDMS slabs and the resulting height when bonded together. (B) Photograph of two actual devices, filled with red dye for visualization and a EU 1 cent coin for scale. (C) Schematic of the experimental set-up. Spheroids enter the PDMS chip at the inlet reservoir, which is a pipette cone, after which they are aspirated with hydrostatic pressure by changing the height $\Delta h$ of the outlet vial that is mounted to a vertical rod. The experiment is captured with an inverted microscope using a 5x objective. (D) Brightfield top view image of the aspiration pockets loaded with aspirated spheroids, scale bar 200 \textmu m.} 
\end{figure}

The master wafer was created using standard soft lithography at the Kavli Nanolab Delft. The design is available at https://github.com/RubenBoot/HighThroughput\_Spheroid\_MPA. The multi-layer design contains features with different heights, so had to be created in two separate photolithography steps using a \textmu MLA laserwriter (Heidelberg Instruments). The final chip was designed as a combination of two slabs of PDMS, one slab with the aspiration channels 50 \textmu m in height plus the top half of the aspiration pockets \textpm 150 \textmu m in height (see Fig. \ref{Fig1}A, PDMS slab 1), and the other slab containing the bottom half of the aspiration pockets (PDMS slab 2). The molds for both these slabs were fabricated together on one silicon wafer. To obtain this, the first step of the design was created by spinning SU-8 3050 to an average thickness of 50 \textmu m and soft baking it at 95 \degree C for 15 minutes, after which the laserwriter wrote the first layer. The wafer was post baked at 65 \degree C for 1 minute, then at 95 \degree C for 5 minutes and developed in SU-8 developer. The second layer was created with SU-8 2050 and spun to an average thickness of 150 \textmu m. It is important to note that the thickness was not equal over the whole wafer, as the resist covered both the first half of the design (consisting of the micropipette channel and part of the aspiration pocket) and the empty place where the second half of the aspiration pockets will be written. Therefore, one half of the aspiration pockets (see Fig. \ref{Fig1}A, PDMS slab 2) resulted in a thickness of 150\textpm 2 \textmu m, while the other half containing half of the pocket plus the aspiration channel (PDMS slab 1) had a different thickness of 175\textpm 25 \textmu m. After spinning, the wafer was soft baked at 65 \degree C for 5 minutes and then at 95 \degree C for 30 minutes. The laserwriter wrote the second part of the design, after which the wafer was post baked at 65 \degree C for 5 minutes, at 95 \degree C for 12 minutes and then developed. The master wafer was coated with trichloro(1H,1H,2H,2H-perfluorooctyl)silane to allow for easy demolding. PDMS chips were created using Sylgard 184 at a curing agent ratio of 10:1. Individual chips were cut and holes were punched in only one slab of the design for the later introduction of tubing. Both halves of the design were plasma cleaned (Harrick Plasma) for two and a half minutes to facilitate bonding, after which one half was slightly wetted with a droplet of distilled water to allow for better alignment. The two halves were put together and aligned using the alignment arrows at the border of the design and an optical microscope (ZEISS Primovert) to check the alignment (Fig. S2\dag). Finally, the chip was kept in the oven at 65 \degree C to bond overnight. Afterwards, chips could be stored indefinitely and used on the desired day for the experiment.  
\subsection*{Data acquisition}
Before each experiment, the chip was filled with 1\% Pluronic\textsuperscript{\textregistered} F127 (Sigma) solution and left at room temperature to prevent cell adhesion to the PDMS walls. After 45 minutes, the pluronic solution was flushed from the chip using the culture media that matched the cell line used in the experiment. For this, vials with cell-free culture media were connected to the inlet and outlet of the microfluidic chip with PTFE 008T16-030-200 tubing (Diba Industries, inner diameter 0.3 mm, outer diameter 1.6 mm) and a pressure was applied to the media using an MFCS-EZ pressure controller (Fluigent). Once all the pluronic solution, PDMS debris particles and possible air bubbles were flushed out, the tube connected to the inlet was gently unplugged from the chip and a loading reservoir, being a shortened 1 ml pipette tip cut with a scalpel, was plugged into the inlet. Then, the vial connected to the outlet was disconnected from the pressure controller and mounted to a vertical rod, to be able to exert a precise hydrostatic aspiration pressure when lowering the vial compared to the height of the reservoir. Slightly lowering the vial induced a minor flow in the chip towards the aspiration pockets, after which 20 \textmu l of spheroid suspension was pipetted into the reservoir. Spheroids were guided by the flow and entered the aspiration pockets, after which the outlet tube was brought back to the height where no flow is present. For loading, the inlet hole needed to be punched close enough to the aspiration pockets. Otherwise, the volume of space in the loading bay of the chip would be too large and could not induce a high enough flow velocity to sweep the spheroids into the pockets. Spheroids then sedimented to the bottom instead and remained immobile. Once ready to perform the experiment, the vial was manually lowered to the chosen aspiration pressure, thereby inducing spheroid tongue aspiration.

Brightfield images of spheroid tongues entering the aspiration channels were captured on an inverted fluorescence microscope (Zeiss Axio-Observer) every 5 seconds for a total of 5 minutes using a 5x/NA 0.16 air objective and ORCA  Flash 4.0 V2 (Hamamatsu) digital camera with a resolution of 2048x2048 px\textsuperscript{2}. We ensured that the full aspiration curve of the tongue was captured by starting the image acquisition before lowering the outlet tube (Movie S1\dag). At the end of the experiment, spheroids were pushed out of the pockets and flown back towards the inlet by raising the outlet vial above the reservoir. There, they were removed from the chip by pipetting them up through the reservoir. This way, new spheroids could be inserted and a new measurement started with the same chip. All experiments were conducted at 37 $\degree$C and 5$\%$ \ce{CO2} using a stage top incubator (ibidi). Chips were used for 4-5 successive experiments on average, and were always discarded after the final experiment of the day.

\subsection*{COMSOL simulations on pressure distribution in the chip}
The design of the device contains 8 parallel pockets, and thus pressure will redistribute once spheroids start clogging the flow in pockets. To examine the influence of this effect, the pressure distribution was computationally modeled in the 3D design of the device using the finite elements modeling software COMSOL Multiphysics 5.6. Considering the fluid flow to be laminar and following the Navier-Stokes equation,\cite{constantin2020navier} the pressure distribution was modeled for two different cases: (1) all pockets are open, or (2) all pockets are clogged, except for one where fluid still flows through the aspiration channel. 

The Hagen-Poiseuille equation tells us that the pressure drop over a tubular channel with laminar flow scales with the length of the channel and the inverse of the channel radius to the fourth power.\cite{Liu2016a} As the design of our device has much larger dimensions than typical microfluidic chips, the cross-sectional area of the tubing connected to the chip now has the same order of magnitude as the cross-sectional area of the device. Therefore, the hydrodynamic resistance across the tubing is considerable and non-negligible as long as a flow is present in the device. It is therefore important to realize that the device is only able to accurately perform spheroid aspiration at a single defined step pressure once \textit{all} pockets are filled with a spheroid, thus blocking the flow.  

To examine what pressure spheroids experience in pockets when not all are filled yet, case (1) and (2) were modeled for the design of the device including a 60 cm long rectangular channel with a 300x300 \textmu m cross section that mimics the outlet tube and corresponding pressure drop. The same average length of tubing was used during our experiments. Boundary conditions of 700 Pa at the inlet and 0 Pa at the outlet were installed, similar to lowering the outlet tube with 7 cmH\textsubscript{2}O. These simulations generated the pressure distribution and corresponding fluid flow profile in the device.  

\subsection*{High-throughput analysis of spheroid tongue aspiration}
\begin{figure}[h]
\centering
\includegraphics[height=11.8cm]{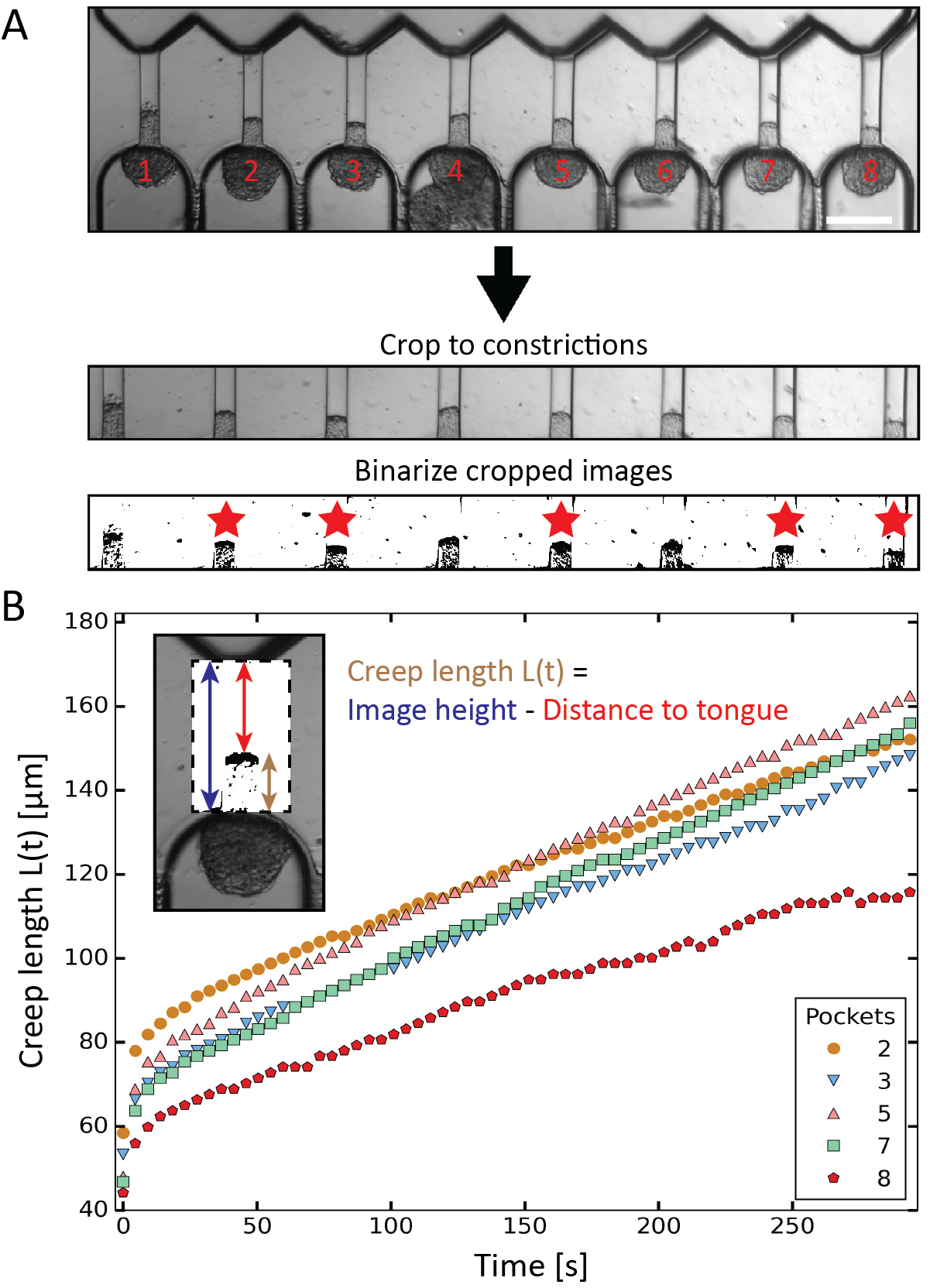}
\caption{\label{Fig2} Automated data analysis using a custom-written Python script. (A) Preparation workflow before running the script. First, the 8 pocket-time series (scale bar 200 \textmu m) is cropped to the constrictions with spheroid tongues, where the bottom of the cropped images is aligned with the precise start of the aspiration channels. Next, the cropped time series is converted to binary images by setting a threshold using ImageJ. Tongues are analyzed if they have a smooth thresholded edge and belong to spheroids that remain round and retain a constant volume before and during aspiration, here indicated by a red star. (B) Plot of the creep length of the 5 selected spheroids (pocket numbers, see (A)) as a function of time. The inset shows how the tongue length is calculated by subtracting the distance from the top of the cropped image to the tongue edge from the total image height.}
\end{figure}

The creep length of the spheroid tongues into the aspiration channels over time was extracted from the experimental images using Fiji (https://imagej.net/software/fiji/) and a custom-written Python script (which is available at: https://github.com/RubenBoot/HighThroughput\_Spheroid\_MPA
/blob/main/SpheroidAspiration\_AnalysisScript.py). First, the brightfield aspiration time-lapse images were rotated to make the aspiration channels vertically oriented (with the tongue creeping upward over time), and then cropped using ZEN software (Zeiss). The cropped region captured the whole aspirated spheroid tongue and aligned the beginning of the channels with the bottom of the cropped images. The cropped time-lapse was saved as a JPEG stack, and converted to binary images using a threshold in Fiji (see bottom two frames in Fig. \ref{Fig2}A). The threshold value was chosen manually to obtain a clear contrast between the aspirated tongue edge and the surrounding empty channel. This binary stack was then imported in the Python script. Using the Fiji interface, the x-coordinates of pixels along the horizontal line were manually inserted in the Python script to indicate where the middle of all 8 aspiration channels was located. The script was set up to find the edge of the aspiration tongue by checking the binary value of every pixel on the vertical line along these coordinates (from top to bottom) and recording the y-coordinate corresponding to the tongue edge. This analyses was repeated for all images in the stack, returning the set of y-coordinates for all 8 channels and for each time step.
Y-coordinates (pixels) were converted into creep lengths (\textmu m) by subtracting the y-coordinate (which is the distance from the top of the image to the tongue edge) from the total height of the image and multiplying this result with the pixel distance. All results were collected in a Microsoft Excel file alongside the time step per image. 
\subsection*{Statistical analysis}

Statistical analysis was performed using Python and Microsoft Excel. Student \textit{t}-tests were executed using the TTEST function in Excel and \textit{p} values below 0.05 were considered to be significant. Python was used to do standard error propagation calculations on the fitting parameters to obtain error values. The average human error in the aspiration pressure when manually lowering the outlet vial is estimated to be 20 Pa ($\sim$ 0.2 cm\ce{H2O}), and the error in dimensions of the aspiration channel is around 5 \textmu m (measured with a Dektak stylus profiler). The error bars in the figures display the standard error of the mean unless indicated otherwise, and are always based on at least two independent experiments with separately prepared chips.

\section*{Results and discussion}
\subsection*{High-throughput data extraction}

In order to measure mechanics of cell spheroids in a high-throughput manner, we designed a microfluidic device to parallelize micropipette aspiration. The design consists of 8 parallel aspiration pockets that are 250 \textmu m wide and 325\textpm 27 \textmu m tall, in order to be able to contain a single spheroid with a maximum diameter of 250 \textmu m. Each pocket connects to an aspiration channel that is 50 \textmu m wide and 50\textpm 2.5 \textmu m tall, chosen to be similar in size to pipette diameters used in previous glass micropipette spheroid aspiration studies (Fig. \ref{Fig1}A-B) .\cite{Dufour2010,Yousafzai2022} We decided to implement 8 pockets as it was the maximum number of pockets that fit in the field of view of the 5x microscope objective. To mimic traditional micropipette aspiration as accurately as possible, the multilayer wafer mold was designed to create symmetric aspiration pockets with aspiration channels positioned at the \textit{centerline} of the pockets. This way, spheroids were raised from the device bottom during aspiration (similar to the single cell aspiration device created by Lee \textit{et al.}).\cite{Lee2015} Two PDMS halves from the device mold, one side with half the pocket and the other with the other half of the pocket plus the aspiration channel, were aligned under a microscope using alignment arrows incorporated in the design (see ESI\dag~ Fig. S2) and bonded in the oven overnight.

The chip was flushed with media before starting an experiment. Then, a small volume of spheroid suspension ($\sim$ 20 \textmu L) was added to the inlet reservoir. The spheroids were moved into the pockets through the flow induced by lowering a media reservoir connected to the outlet and mounted to a vertical rod, thereby exerting a precise hydrostatic pressure gradient (Fig. \ref{Fig1}C). Once a spheroid arrived in the pocket, it blocked the flow through the aspiration channel, thus preventing other spheroids from entering the same pocket. Induced pressure gradients, by lowering the outlet media reservoir, were at first kept low enough to ensure that spheroids did not deform in the aspiration channels yet. Only when all pockets were loaded with spheroids was the outlet reservoir lowered to a level that induced the chosen step pressure for aspiration. Spheroid tongues started creeping in the aspiration channels as they were subjected to the pressure difference between the atmospheric pressure at the inlet reservoir and the hydrostatic pressure exerted by the outlet media tube (Fig. \ref{Fig1}D). In this way, spheroids experienced the aspiration force in a similar manner as for traditional micropipette aspiration, where spheroids are kept at atmospheric pressure and aspirated by applying an underpressure in a glass capillary. 

We developed a custom-made Python script to analyze the spheroid tongue deformations in an efficient and fast manner. Before running the script, brightfield time-lapse image series were rotated, cropped and changed into binary images to focus on the creeping tongues in the channels (Fig. \ref{Fig2}A). To determine the creep length $L(t)$, the script derived the distance from the top of the binarized image to the spheroid tongue edge, and subtracted this value from the total image height. The Python script thus extracted the creep lengths for the 8 aspiration channels and each step in time (Fig. \ref{Fig2}B). We excluded spheroids that were not round before- or did not remain constant in volume during aspiration.

\subsection*{Pressure distribution in the chip}

In contrast to microfluidic single cell aspiration chips, where channels have smaller dimensions than the used tubing,\cite{Davidson2019,Lee2015} the required cross-section for channels to flow undeformed spheroids in is as large as the tubing (with a diameter of 300 \textmu m). 
\begin{figure}[h]
\centering
\includegraphics[height=12.8cm]{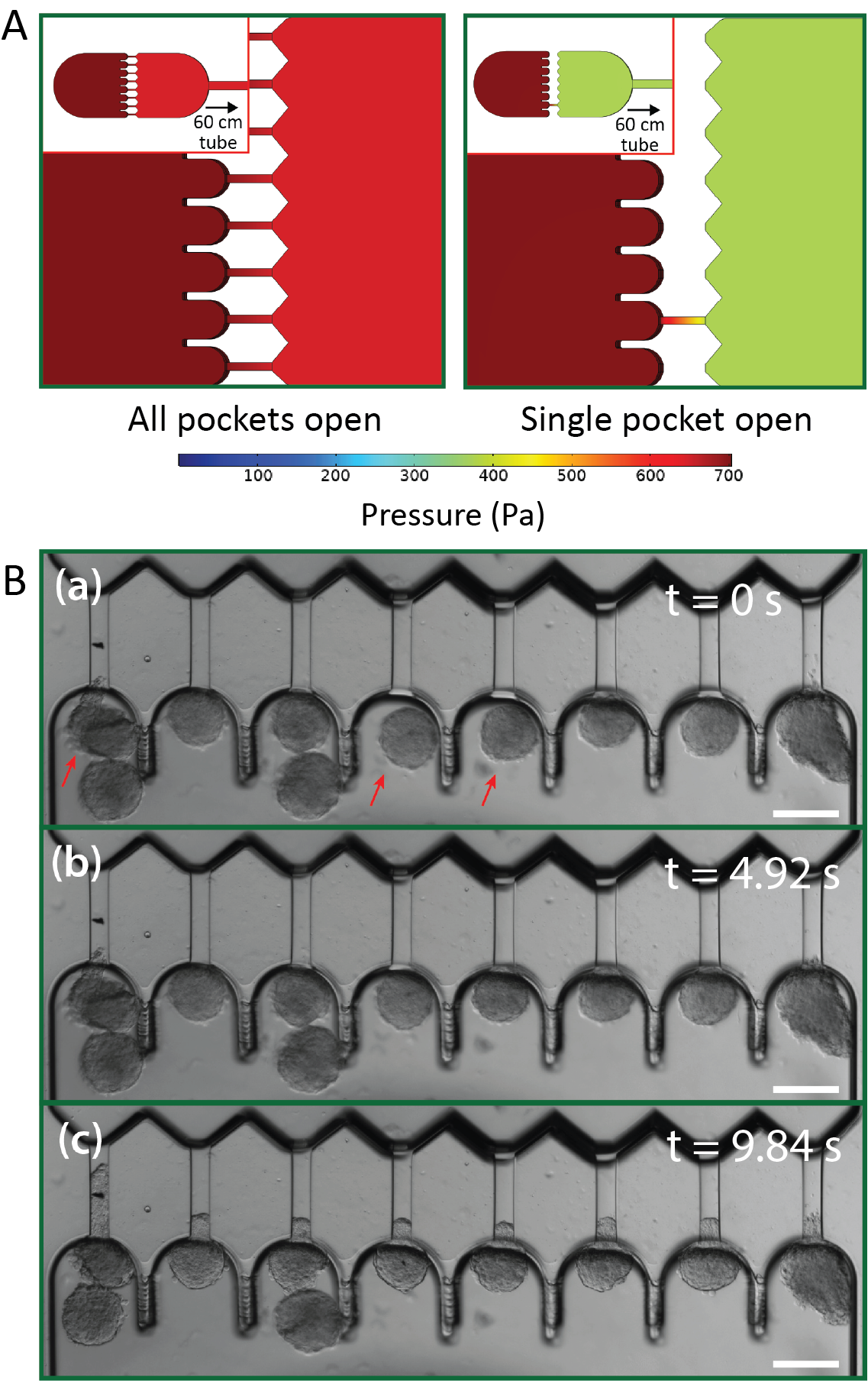}
\caption{\label{Fig3} Pressure distribution across aspiration pockets. (A) A 3D numerical simulation of the pressure distribution assuming the device is connected to a 60 cm long straight rectangular channel mimicking the outlet tube. A pressure gradient of 700 Pa is simulated for two different configurations: all pockets are open (left), and all pockets but one are clogged by spheroids (right). (B) Brightfield images from an HEK cell spheroid aspiration experiment, demonstrating how simultaneous aspiration starts as soon as all channels are clogged. At first (a), no pressure gradient is present yet and three pockets remain open, indicated by red arrows. Five seconds later (b), a pressure gradient has been induced and spheroids flow into the pockets but do not clog everything yet. After five more seconds (c), all spheroids have clogged the pockets, thereby blocking further flow and ramping up the pressure. Spheroids now experience the full induced hydrostatic pressure gradient as evident from the tongues all creeping simultaneously into the aspiration channels. Scale bar 200 \textmu m.}
\end{figure}
Therefore, the pressure drop over the tubing is non-negligible in comparison to the pressure drop across the chip. This is a considerable problem, as slight changes in the length of the tubing may cause significant changes in the experienced pressure drop in the microfluidic device. To avoid these pressure drop effects, we designed the device in such a way that the entire flow is stopped once all pockets are filled with spheroids and the pressure gradient is fully defined by the hydrostatic pressure. However, spheroids that are loaded into the device and swept along by the induced flow will reach the aspiration pockets one at a time. Therefore, pressures will redistribute in the device for each spheroid that clogs an aspiration pocket and stops the flow through the adjacent aspiration channel. This might cause a problem, as the spheroids that already arrived in the pockets may experience an aspiration pressure that increases over time when additional pockets fill up with spheroids. Instead, all spheroids should experience a single step pressure to allow correct data analysis. 

To examine the effect of pressure redistribution, we performed 3D COMSOL simulations for two different cases: (1) all pockets are open, and (2) all pockets are clogged by spheroids except for one. A 60 cm long rectangular channel with a 300x300 \textmu m cross section was incorporated into the 3D design to mimic the outlet tube, as it induces a considerable pressure drop, and a pressure gradient of 700 Pa was simulated across this total geometry. The model showed how the pressure gradient over the aspiration channel increases from $\sim$50 Pa when all pockets are open to $\sim$300 Pa when all but one are clogged (Fig. \ref{Fig3}A). It is important to note that these values are dependent on the chosen boundary conditions, and in this context simply serve to estimate the extent of this effect. 

We concluded from these simulations that it is important to ensure that spheroids arrive at the pockets at approximately the same moment in time. Otherwise, spheroids might experience a pressure that starts aspiration when \textit{almost} all pockets are clogged but jumps once the final pocket clogs. In experiments, we discovered that spheroids all start aspirating at the same time as long as we punch the inlet hole close to the pockets and add sufficient spheroids to the inlet reservoir (Fig. \ref{Fig3}B). To circumvent the effect of pressure redistribution, spheroids were slowly flown towards the pockets by inducing a minor pressure gradient ($\sim$150 Pa) and were halted before the pockets by bringing the height of the outlet vial back to the starting point with no flow. Then, the final pressure gradient was induced by lowering the outlet vial again, and all spheroids that were floating near the pockets experienced the flow and filled up the remaining unclogged pockets at the same time. This ramped up the pressure, and the spheroids experienced the full pressure gradient, thus starting the measurement at the same time for all pockets. 
\begin{figure*}
\centering
\includegraphics[height=5.4cm]{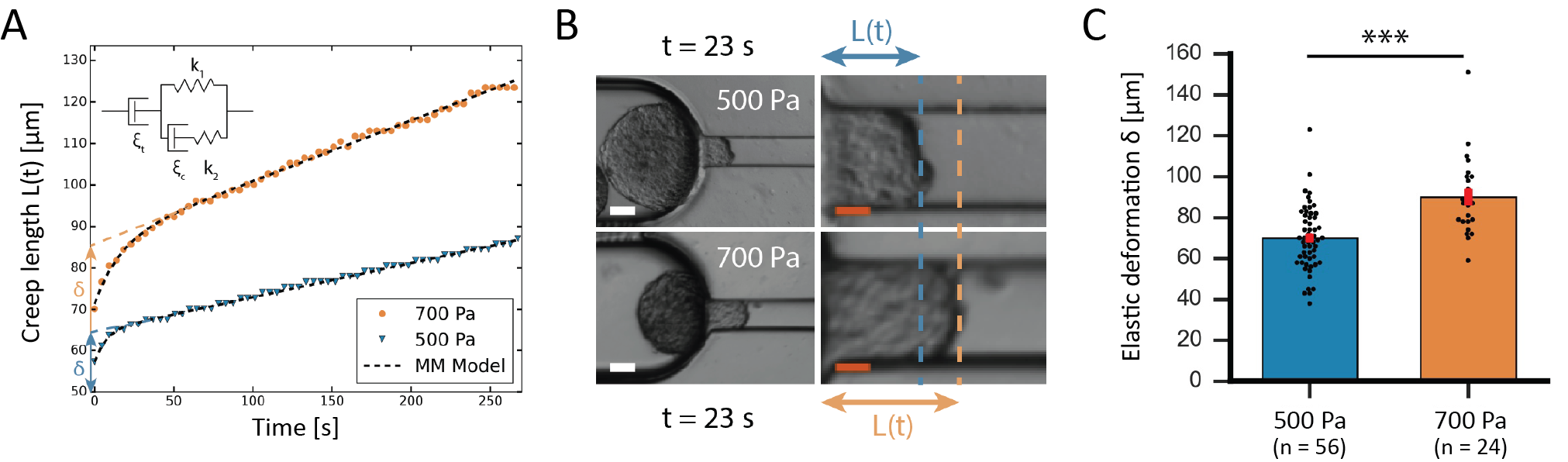}
\caption{\label{Fig4} Validation of sensitivity of the device to determine spheroid mechanical properties. (A) Comparison plot of the creep length versus time for one HEK cell spheroid aspirated at 700 Pa (orange) and another at 500 Pa (blue). The creep curves are fitted with the modified Maxwell model (black dashed lines), which is explained in the inset. The fast elastic deformation $\delta$ equals the intercept of the linear viscous flow with the y-axis, and is shown to be larger when aspirating spheroids at larger pressures. (B) Brightfield images of two separate aspiration experiments for HEK cell spheroids, showing the creep length \textit{L(t)} after 23 s of aspiration at 500 Pa (top) or 700 Pa (bottom), with snapshots focusing on the aspiration pocket (left, scale bar 50 \textmu m) and zoomed in on the aspiration channel (right, scale bar 20 \textmu m). (C) Histograms comparing the average elastic deformation $\delta$ for HEK cell spheroids aspirated at different pressures (500 Pa and 700 Pa). ***, p < 0.001 with n = 56 and 24 for 500 Pa and 700 Pa respectively. Error bars are SEM.}
\end{figure*}
\subsection*{Validation of sensitivity and reproducibility when working at different aspiration pressures}

To confirm the sensitivity and reproducibility of our device, we measured the deformability of HEK293T spheroids at two slightly different aspiration pressures of 500 and 700 Pa. The obtained creep data alongside visual confirmation showed that the aspirated spheroids displayed an initial elastic deformation followed by a viscous response (Fig. \ref{Fig4}A). The tissue relaxation time $\tau$ demarcates these two regimes and is given by $\tau$ = $\eta$/$E$, with $\eta$ being the viscosity and $E$ the elastic modulus of the spheroid. We fitted this viscoelastic response to the modified Maxwell model, following previous studies on spheroid micropipette aspiration.\cite{Dufour2010,Yousafzai2022,Guevorkian2011} The model consists of four elements (shown in the inset of Fig. \ref{Fig4}A): a dashpot $\xi_t$ in series with a modified Kelvin-Voigt element, which consists of a spring $k_1$ in parallel with a dashpot $\xi_c$ and spring $k_2$ in series. The creep length $L(t)$ in the context of this model is given by:
\begin{equation}\label{Lequation}
  L(t) = \frac{f}{k_1}(1-\frac{k_2}{k_1+k_2}e^{\frac{-t}{\tau_c}})+\frac{f}{\xi_t} t
\end{equation}
where $f$ is the aspiration force, $k_1$ is the spring constant for the elasticity of the spheroid, $k_2$ relates to the initial increase in $L(t)$, $\xi_t$ represents the viscous dissipation of the flowing tongue and $\tau_c$ is the rising time of the elastic deformation $\delta$. Here, $\tau_c$ = $\frac{\xi_c (k_1 + k_2)}{k_1 k_2}$ with $\xi_c$ being a friction coefficient related to the rising time. Hence, the modified Maxwell model has four fitting parameters: $\delta$ = $f$/$k_1$, $\dot{L_\infty}$ = $f$/$\xi_t$ (being the flow velocity at long timescales), $\beta$ = $k_2$/($k_1$ + $k_2$) and $\tau_c$.

As shown in Fig. \ref{Fig4}B, the creep length $L(t)$ of the spheroid tongue was smaller for measurements at 500 Pa in comparison to 700 Pa, demonstrating the sensitivity of our device. However, creep length can only be compared for a precise point in time. To quantify differences between the entire measurements, we compared the average elastic deformation $\delta$ by fitting Eq. (\ref{Lequation}) to the creep data. The average deformation $\delta$ was significantly smaller for measurements performed at 500 Pa ($\delta$ = 70 $\pm$ 2 \textmu m) in comparison to 700 Pa ($\delta$ = 90 $\pm$ 4 \textmu m), as one would expect when exerting a smaller force (Fig. \ref{Fig4}C). This demonstrates how the device is sensitive enough to work at small differences in pressure and create reproducible deformations in agreement with theoretical expectations. 

\subsection*{Measurements on spheroids with known mechanical properties}
\begin{figure}[h]
\centering
\includegraphics[height=8.3cm]{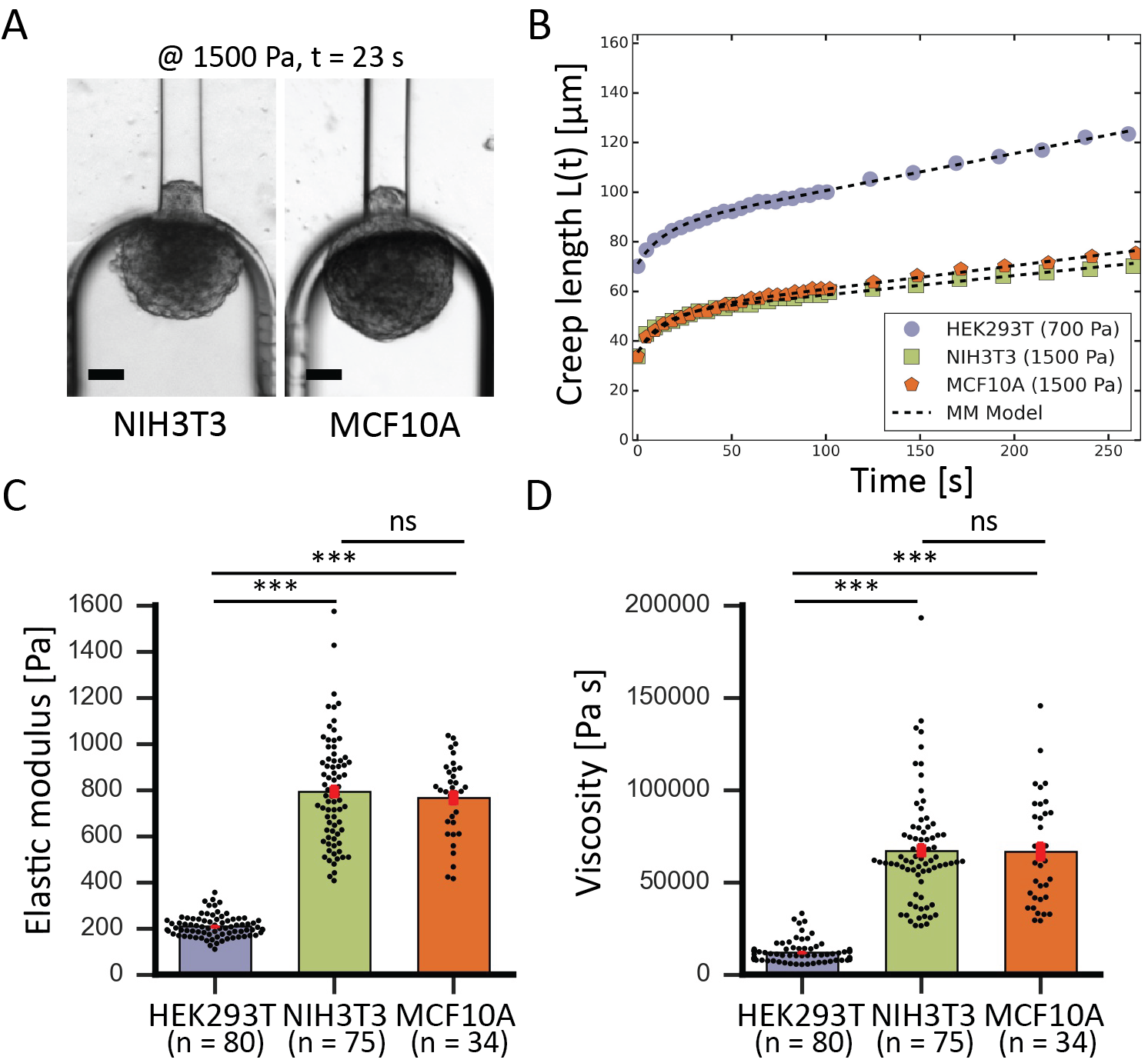}
\caption{\label{Fig5} High-throughput comparison of spheroid mechanics for different cell lines. (A) Brightfield snapshots after 23 s of aspiration at 1500 Pa of a NIH3T3 (left) and MCF10A (right) spheroid, scale bars 50 \textmu m. (B) Comparison plot of creep length versus time for a HEK293T spheroid (purple) aspirated at 700 Pa, and a NIH3T3 (green) and MCF10A (orange) spheroid aspirated at 1500 Pa. The creep curves are fitted with the modified Maxwell model (black dashed lines). (C) Average elastic moduli and (D) average viscosity measured for the three different cell lines. ***, p < 0.001 and ns is nonsignificant with n = 80, 75 and 34 for HEK293T, NIH3T3 and MCF10A respectively. Error bars are SEM.}
\end{figure}

To demonstrate the high-throughput mechanical phenotyping capabilities of our device, we measured the mechanical properties of three different cell spheroid models that have been probed in previous studies.\cite{Efremov2021} In addition to our measurements performed with HEK293T cell spheroids, we studied two stiffer spheroid models made of either NIH3T3 fibroblasts or MCF10A cells. While the HEK293T cell spheroids were probed at 500 and 700 Pa, the two stiffer spheroid models were aspirated at a higher pressure of 1500 Pa as lower pressures would induce a slower deformation and require a longer time scale to analyze the full viscoelastic response (Fig. \ref{Fig5}A).  

In traditional micropipette aspiration, the aspiration force $f$ of the pipette when considering spheroid volume conservation is given by:
\begin{equation}\label{aspforce}
f=\pi R_{p}^2 \Delta P,
\end{equation}
where $R_p$ is the radius of the pipette and $\Delta P$ the applied underpressure in the pipette.\cite{Hochmuth2000} Previous studies on micropipette aspiration of spheroids have pointed out that the actual pressure exerted on the spheroid equals the applied underpressure $\Delta P$ minus a critical pressure $\Delta P_c$ at which aspiration of the spheroid occurs. When aspirating at a pressure lower than $\Delta P_c$, the spheroid will not deform due to its inherent tissue surface tension $\gamma$.\cite{Dufour2010,Yousafzai2022} 

To calculate this critical pressure, separate measurements of the retraction of the spheroid tongues are required when the pressure is brought back to zero (see ESI\dag). In contrast with traditional glass micropipette aspiration, where spheroids remain stuck in the pipette and the tongue slowly retracts, on our device spheroid tongues retracted so fast that it was impossible to measure a retraction curve when the pressure was brought back to zero (see Movie S2\dag). This effect was consistent for all three cell lines. We therefore investigated the effect of accounting for $\Delta P_c$ through traditional micropipette measurements on HEK293T spheroids (measuring both aspiration and retraction, see ESI\dag). We found that $\Delta P_c$ ranged between 50 and 150 Pa (when aspirating at 500 Pa), which changes parameter values obtained by fitting aspiration curves by maximally 10-30 \% (Fig. S3 and Table S1\dag). Additionally, we saw that the spheroids deformed differently in the glass micropipettes in comparison to our device, displaying a lower elastic deformation $\delta$ and flow velocity $\dot{L_\infty}$. One possible explanation for the discrepancy between our device and traditional MPA is that the chip has square aspiration channels instead of round capillaries, potentially influencing the creep of the tongue due to the different geometry or due to wall friction (PDMS instead of glass). Moreover, the pockets on our chip have rounded walls which the spheroids potentially push back on during relaxation, making them move slightly backwards in the pocket during retraction. We conclude that measurements with our microfluidic device slightly deviate from glass micropipette aspiration measurements. However, our device has the unique benefit of providing measurements at high throughput, thus allowing systematic comparisons between different cell types.

The creep curves showed that the two stiffer spheroid models made of NIH3T3 or MCF10A cells had a lower initial elastic deformation $\delta$ and slower viscous flow $\dot{L_\infty}$ than the more deformable HEK293T spheroids (Fig. \ref{Fig5}B). All curves were fitted with the modified Maxwell model to extract the relevant mechanical parameters. The first term in Eq. (\ref{Lequation}) characterizes the elastic regime, with $k_1$ = $\pi$$R_{p}$$E$, while the second term represents the flow at constant velocity $\dot{L_\infty}$ for longer timescales, with $\xi_t$ = 3$\pi^2\eta R_{p}$. As our microfluidic device, unlike traditional glass micropipettes, does not present cylindrical constrictions, a correction for rectangular constrictions needs to be implemented in regards to $R_p$.\cite{Son2007,Davidson2019} 
The effective channel radius $R_{eff}$ is now given by:
\begin{equation}\label{Reff}
R_{eff}^4=\frac{2}{3\pi}\frac{W\times H^3}{(1+\frac{H}{W})^2\times f^{*}},
\end{equation}
with $W$ and $H$ being the width and height of the rectangular constriction and $f^*$ being a function of the aspect ratio ($H$/$W$), given by Son.\cite{Son2007}

When fitting obtained creep data with Eq. (\ref{Lequation}) and plugging in Eq. (\ref{aspforce}, the elastic modulus $E$ is derived as:
\begin{equation}\label{elasticmodulus}
E = \frac{R_{eff} \Delta P}{\delta},
\end{equation} 
and the viscosity $\eta$ as:
\begin{equation}\label{viscosity}
\eta = \frac{1}{3\pi\dot{L_\infty}}R_{eff} \Delta P.
\end{equation} 
The average elastic moduli were almost four times as large for both NIH3T3 and MCF10A cell spheroids in comparison to the softer HEK293T cell spheroids, but were not significantly different from each other (Fig. \ref{Fig5}C). Similarly, both models had a viscosity almost six times larger than for HEK293T spheroids but again were not significantly different from each other (Fig. \ref{Fig5}D).

These obtained values are consistent with values measured in previous studies (Table \ref{Table1}). For HEK293T cell spheroids, we measured an average elastic modulus of $\sim$210 Pa which agrees with parallel-plate compression on HEK293 cell spheroids measuring a range of 200-400 Pa.\cite{Koudan2020} Our NIH3T3 spheroids had an average modulus of $\sim$780 Pa, which falls within the range of 500-3500 Pa measured with colloidal probe AFM.\cite{Jorgenson2017} However, the MCF10A spheroids displayed an average modulus of $\sim$770 Pa which is just below the range of 1250$\pm$320 Pa measured by squeezing MCF10A spheroids with microtweezers.\cite{Jaiswal2017} This discrepancy could be explained by the fact that this range was determined for only 6 spheroids in the microtweezer study, or potentially squeezing might induce different deformation behavior compared to aspiration. Altogether, these results demonstrate that our device is well capable
\begin{table}[h]
\small
  \caption{\ Spheroid mechanical parameters for different cell lines, derived using the modified Maxwell model and performing a least squares regression of the experimental creep curves}
  \label{Table1}
  \begin{tabular*}{0.48\textwidth}{@{\extracolsep{\fill}}llll}
    \hline
    Cell line & \textit{E} (Pa) & $\eta$ (kPa s) & n \\
    \hline
    HEK293T & 210 (\textpm 5) & 12 (\textpm 1) & 80 \\
    NIH3T3 & 780 (\textpm30) & 67 (\textpm 3) & 75\\
    MCF10A & 770 (\textpm 30) & 67 (\textpm 5) & 34\\
    \hline
  \end{tabular*}
\end{table}
to measure the viscoelastic behavior of multicellular spheroids and determine their mechanical properties in agreement with other experimental techniques.

\section*{Outlook and conclusions}
We have developed a microfluidic chip that follows the principles of traditional micropipette aspiration to quantify the viscoelastic response of cell spheroids in a high-throughput manner, making it possible to make statistically meaningful comparisons between different experimental conditions. The chip performs viscoelastic creep measurements as soon as spheroids fill up the 8 parallel aspiration pockets and block further fluid flow. The design of the chip can in principle be adapted to obtain smaller or larger aspiration channels, but this will influence the overall volumetric flow rate through the chip and thus the ease of loading spheroids. Similarly, the number of pockets could be increased, though this would make it more difficult to load them. With the current geometry, our chip is able to obtain reproducible and accurate results and to detect differences upon small changes in pressure. Our results are in agreement with values from previous studies and demonstrate the high-throughput aspect of the chip: it is efficient and easy to use in comparison to traditional micropipette aspiration and other experimental techniques, and provides large amounts of data in a smaller amount of time. Therefore, the microfluidic device presented here is a suitable technique to investigate the mechanics of a wide range of tissues, from embryonic to tumor, to provide mechanistic insights in important physiological processes such as tissue remodeling and cancer metastasis.   

\section*{Conflicts of interest}
There are no conflicts to declare.

\section*{Acknowledgements}
R.C.B. and P.E.B. gratefully acknowledge funding from the European Research Council (ERC) under the European Union’s Horizon 2020 research and innovation program (grant agreement no. 819424). A.R. was supported by an Erasmus+ Traineeship. G.H.K. gratefully acknowledges funding from the VICI project \textit{How cytoskeletal teamwork makes cells strong} (project number VI.C.182.004) which is financed by the Dutch Research Council (NWO). The authors thank Dimphna Meijer for the HEK293T cells, and thank Timon Idema for helpful discussions.




\renewcommand\refname{References}

\bibliography{PhD} 
\bibliographystyle{rsc} 

\newpage
\newcommand{\beginsupplement}{%
        \setcounter{table}{0}
        \renewcommand{\thetable}{S\arabic{table}}%
        \setcounter{figure}{0}
        \renewcommand{\thefigure}{S\arabic{figure}}
        \setcounter{equation}{0}
        \renewcommand{\theequation}{S\arabic{equation}}%
     }
\twocolumn[
    \begin{@twocolumnfalse} 
\section*{Supplementary Information}
\beginsupplement
\subsection*{Glass capillary micropipette aspiration}
HEK293T spheroids were aspirated using pipettes with a diameter of 65$\pm$5 \textmu m, fabricated by pulling borosilicate glass pipettes (Harvard Apparatus, 1 mm OD, 0.5 mm ID) with a laser-based puller (Sutter Instruments Co Mode P-2000) and cutting them with a quartz tile. Cell adhesion to the pipette walls was prevented by incubating the pipettes in 2 mg/mL PolyEthyleneGlycol-PolyLysine (PLL(20)-g[3.5]-PEG(2)/PEG(3.4)-Biotin(20\%), SuSos AG, Dubendorf, Switzerland) in MRB80 solution (80 mM piperazine-N,N’-bis(2-ethanesulfonic acid) (Pipes), pH 6.8, 4 mM MgCl2, 1 mM EGTA [Sigma]) for 30 minutes. Spheroids were suspended in CO2-equilibrated medium and kept in an experimental chamber consisting out of microscopic slides. For this, slides were stuck to a custom-made aluminum spacer of 3 mm thickness using vacuum grease (Beckman Coulter). The pipette was introduced into the chamber, aligned with a spheroid and an aspiration pressure was attained by vertically displacing a water reservoir connected to the pipette using a vertical translational stage (LTS300, Thorlabs). Spheroids were aspirated using a pressure $\Delta P$ = 5 cmH2O and visualized on an inverted microscope (Nikon Eclipse TI) with a 10x air objective. After aspiration, the pressure gradient was removed and the retraction of the tongue was recorded. The creep advancement of the tongue was recorded with an ORCA Flash 4.0 digital camera using a 1s interval for a total of 10 minutes, of which 5 minutes corresponded to aspiration and the next 5 minutes to retraction. Data was obtained for two independent experiments, which were performed at room temperature. As the experimental set-up did not include a heating stage to keep the experimental chambers at the physiological temperature of 37 \degree C, aspiration of spheroids was only performed in the first hour after they came out of the incubator.
The critical pressure $\Delta P_c$ to aspirate the spheroids was derived from
\begin{equation}\label{Supplequation}
  \Delta P_c = \Delta P \frac{\dot{L_R}}{\dot{L_R}+\dot{L_A}},
\end{equation}
where $\dot{L_R}$ and $\dot{L_A}$ are the retraction and aspiration flow rates respectively \cite{Dufour2010,Yousafzai2022}.

\section*{Supplementary Movies}
\beginsupplement
\textbf{Movie 1: Microfluidic multi-channel aspiration experiment to determine spheroid mechanics.}
A brightfield video of HEK293T spheroids being aspirated into the aspiration channels under an applied hydrostatic pressure of $\Delta P$ = 700 Pa for a total duration of 5 minutes. Scale bar is 200 \textmu m. 
\\
\\
\textbf{Movie 2: Spheroids move out of pockets during retraction measurement.}
A brightfield video of NIH3T3 spheroids being aspirated under a hydrostatic pressure of $\Delta P$ = 1500 Pa for the first 10 minutes, after which the pressure gradient is removed and spheroid tongues start retracting. The tongues retract within seconds, making it impossible to record a retraction curve and extract a retraction rate $\dot{L_R}$ to derive a critical pressure $\Delta P_c$. Scale bar is 200 \textmu m.
\\
\\
Movies are available upon request to the corresponding author.

    \end{@twocolumnfalse} 
        ]

\section*{Supplementary Figures}
\beginsupplement
See below.
\begin{figure*}
\centering
\includegraphics[height=12cm]{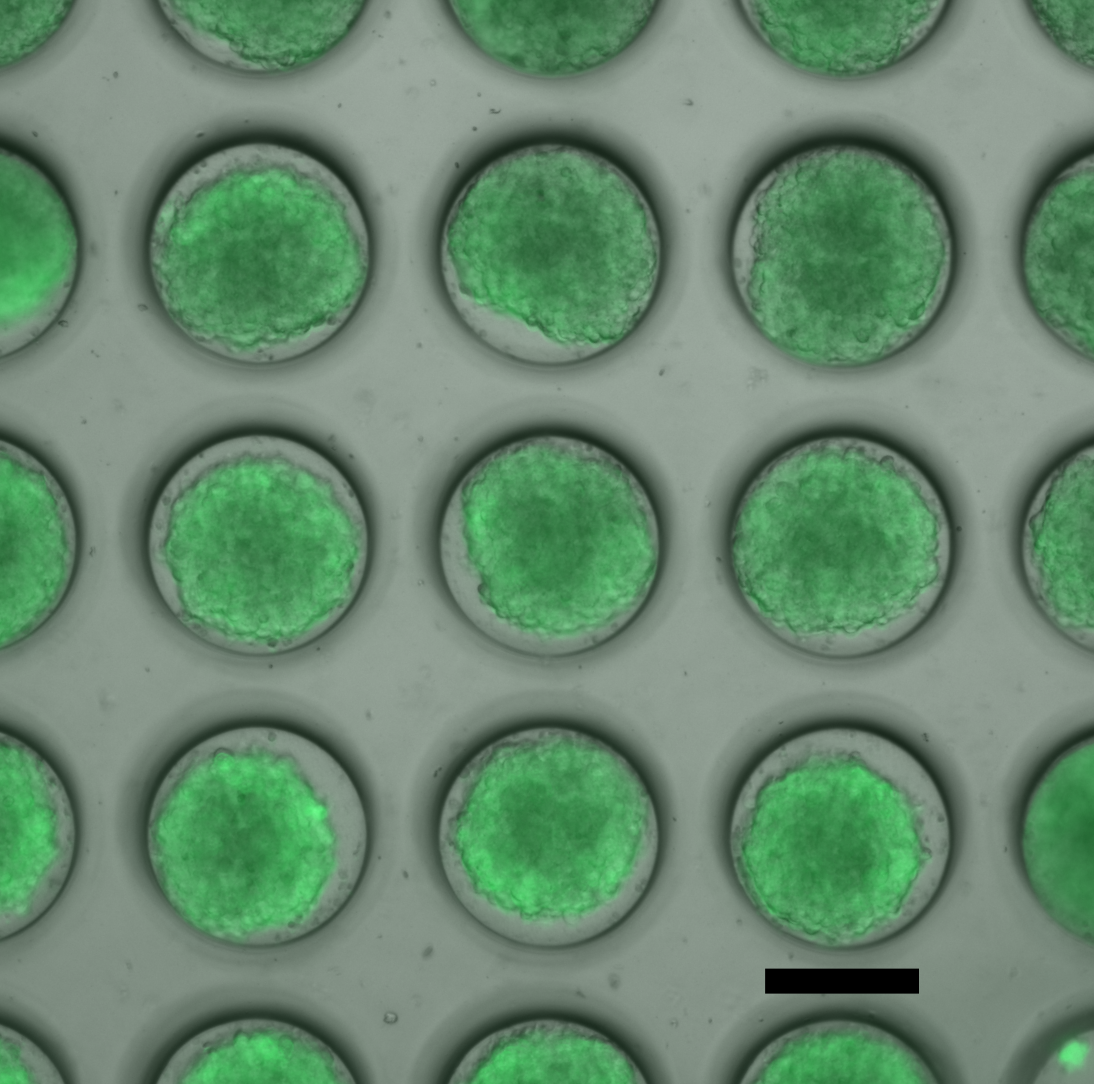}
\caption{Close-up of the microwell array with HEK293T cell spheroids after 3 days of culture. After depositing the cell suspension in the microwell array, cells divide over the wells and aggregate into spheroids to reach a final size that depends on the culturing time, original cell number, and cell type. Here, cells are stained with Calcein AM (AAT Bioquest) for visualization and confirmation of cell viability. The image is an overlay of a brightfield image and a fluorescence image, cropped from an original image taken with an inverted fluorescence microscope (Zeiss Axio-Observer) using a 5x/NA 0.16 air objective. Scale bar 200 \textmu m.}
\end{figure*}
\begin{figure*}
\centering
\includegraphics[height=16cm]{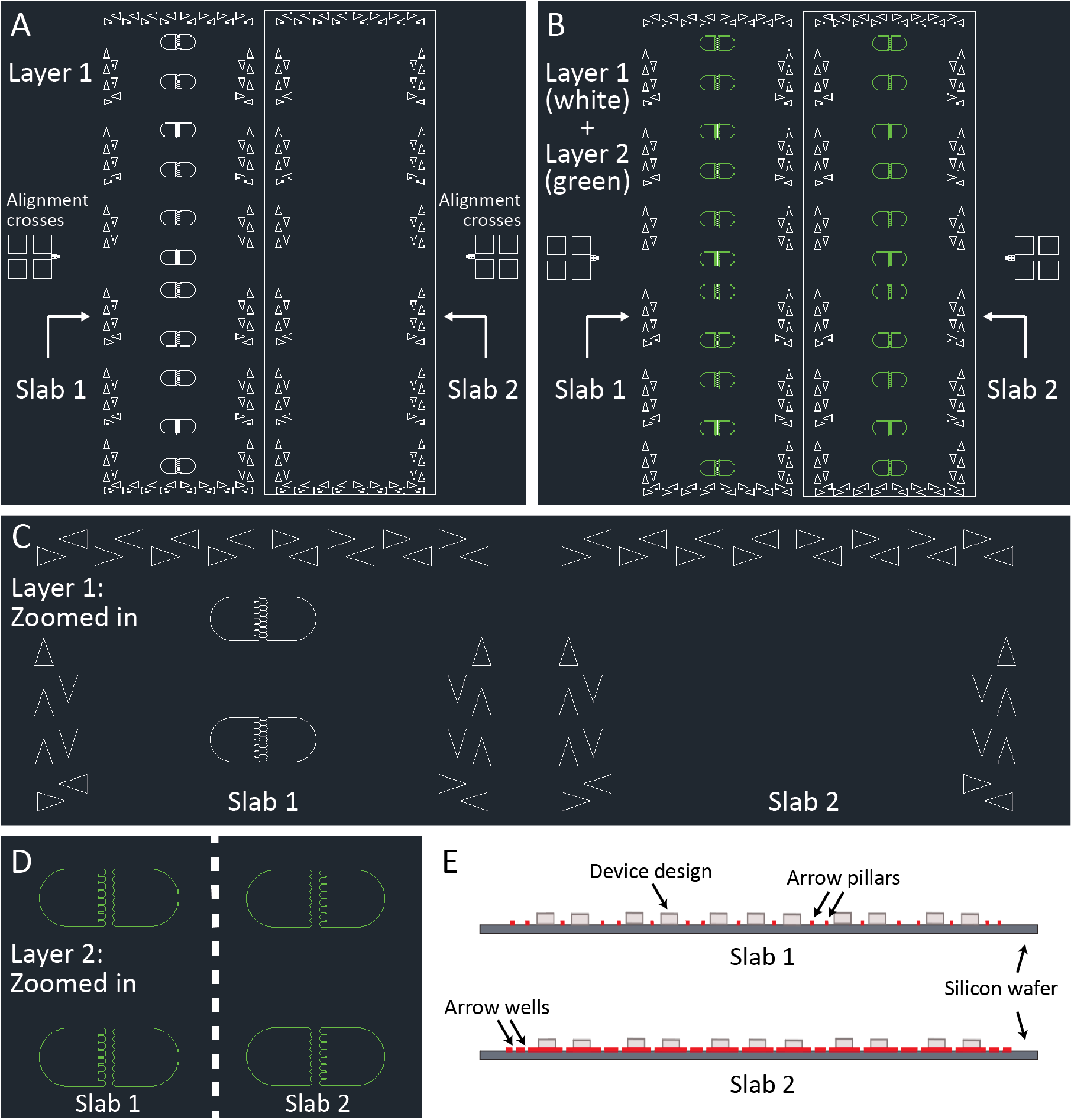}
\caption{Alignment of PDMS slabs using alignment arrows in the design. (A-D) AutoCAD screenshots of the full design that was printed on the silicon wafer using soft lithography. (A) First layer, which was printed to a height of $\pm$50 \textmu m, with the design enframed with alignment arrows. For slab 1, the alignment arrows are written by the laserwriter as pillars, while for slab 2 an entire rectangular surface is written except for empty wells complementary to the arrows of slab 1. Alignment crosses are written to facilitate the alignment of layer 2 with layer 1 in the next step of the writing process. (B) Final developed design, where the second written layer is indicated in green. (C) Zoomed-in section of layer 1, clarifying the placement of alignment arrows. (D) Zoomed-in section of layer 2 (with the middle part cut out) to demonstrate how the right part of the wafer design is mirrored to the left part. This is necessary to create the correct final design, as PDMS slab 2 is turned over when aligning it with PDMS slab 1. (E) Side view of the final developed design on the silicon wafer, which is used for PDMS casting. Slab 1 contains arrow pillars and slab 2 complementary arrow wells (both indicated in red) to facilitate easy alignment before bonding the two slabs in order to create the final device.}
\end{figure*}
\begin{figure*}
\centering
\includegraphics[height=10cm]{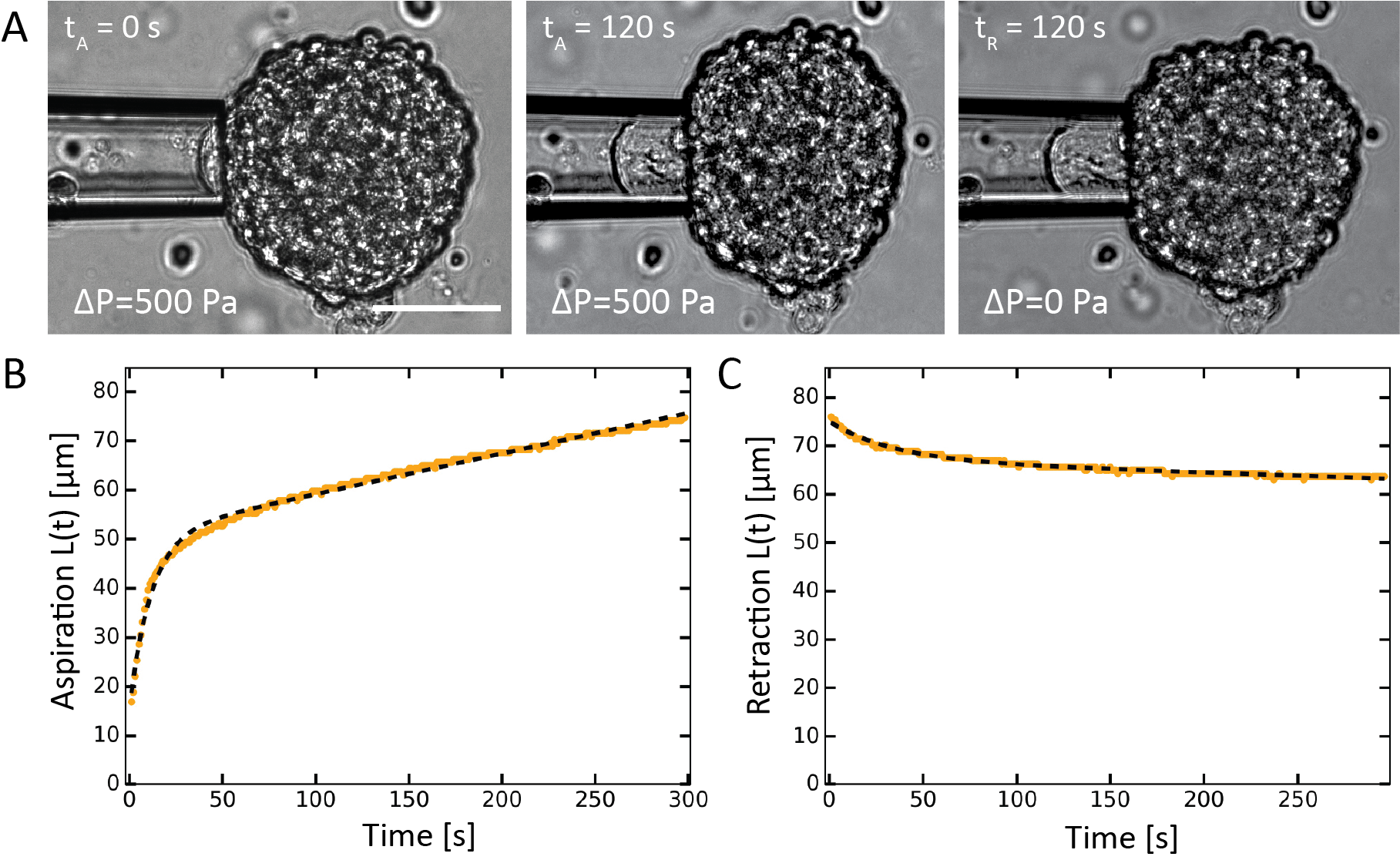}
\caption{Glass micropipette aspiration experiment. (A) Brightfield images of the micropipette aspiration of an HEK293T cell spheroid using a glass micropipette with a radius $R_p$ = 32.5 \textmu m. Left and middle panels: images taken during aspiration at a hydrostatic underpressure of $\Delta$P = 500 Pa, at the beginning of the experiment at $t_A$ = 0 s (left) and after $t_A$ = 2 minutes (middle). Right image: After 5 minutes, the pressure gradient is removed and the retraction is measured, here shown after $t_R$ = 2 minutes (right). Scale bar 100 \textmu m. (B) Creep curve (yellow) of the aspiration of the tongue over time, and (C) creep curve (yellow) for the retraction of the tongue, both fitted with the Modified Maxwell model (black dashed lines).}
\end{figure*}

\section*{Supplementary Tables}
\beginsupplement
See below.

\begin{table*}[h]
\small
  \caption{\ Mechanical characteristics of spheroids with different radii $R_0$ measured by glass micropipette aspiration. The elastic modulus $E$ and viscosity $\eta$ were derived from creep curves fitted to the Modified Maxwell Model, either using just the applied aspiration pressure $\Delta P$ or taking the aspiration pressure corrected for the critical pressure $\Delta P_c$  (from Eq. (S1) in the Supplementary Information), $\Delta P$ – $\Delta P_c$}
  \begin{tabular*}{1\textwidth}{@{\extracolsep{\fill}}lllllll}
    \hline
    Nr. & $R_0$ (\textmu m) & $E_{\Delta P-\Delta Pc}$ (Pa) & $\eta_{\Delta P-\Delta Pc}$ (kPa s) & $E_{\Delta P}$ (Pa) & $\eta_{\Delta P}$ (kPa s) & $\Delta P_c$ (Pa) \\
    \hline
    $1$ & $92$ & $317$ & $30$ & $383$ & $36$ & $86$ \\
    $2$ & $76$ & $237$ & $21$ & $345$ & $31$ & $156$ \\
    $3$ & $99$ & $328$ & $22$ & $419$ & $28$ & $109$ \\
    $4$ & $101$ & $276$ & $18$ & $319$ & $21$ & $68$ \\
    $5$ & $84$ & $281$ & $23$ & $364$ & $30$ & $113$ \\
    $6$ & $87$ & $270$ & $26$ & $374$ & $37$ & $138$ \\
    \hline
  \end{tabular*}
\end{table*}

\end{document}